\documentclass{aa}  

\usepackage{graphicx}
\usepackage{color}
\usepackage{txfonts}

\def\ltsima{$\; \buildrel < \over \sim \;$}
\def\simlt{\lower.5ex\hbox{\ltsima}}
\def\gtsima{$\; \buildrel > \over \sim \;$}
\def\simgt{\lower.5ex\hbox{\gtsima}}
\def\kms{\mbox{km s$^{-1}$}}

\def\cm2{\mbox{$\mbox{cm}^{-2}$}}
\def\cm3{\mbox{$\mbox{cm}^{-3}$}}
\def\h2{\mbox{$_{\mbox{\tiny H2}}$}}

\newcommand{\veck}	{{\bf k}}
\begin{document}

%   \title{The role of molecular filaments in the origin of the stellar initial mass function}
 \title{On the typical width of \textsl{Herschel} filaments}

%   \subtitle{I. Overviewing the $\kappa$-mechanism}

\titlerunning{The typical width of \textsl{Herschel} filaments}

   \author{Ph.~Andr\'e\inst{1}
         \and
          P.~Palmeirim\inst{2}   
          \and
          D.~Arzoumanian\inst{3}
%          \and
%          E.~Schisano\inst{4}          
%          \fnmsep\thanks{Just to show the usage of the elements in the author field}
          }

   \institute{Laboratoire d'Astrophysique (AIM), Universit\'e Paris-Saclay, Universit\'e Paris Cit\'e, CEA, CNRS, AIM, 91191 Gif-sur-Yvette, France %\\
   %CEA/DRF, CNRS, Universit\'e Paris-Saclay, Universit\'e Paris Diderot, Sorbonne Paris Cit\'e, 91191 Gif-sur-Yvette, France\\
              ~~\email{pandre@cea.fr}
                  \and
             Instituto de Astrof\'isica e Ci{\^e}ncias do Espa\c{c}o, Universidade do Porto, CAUP, Rua das Estrelas, PT4150-762 Porto, Portugal\\    
             \email{pedro.palmeirim@astro.up.pt}
         \and
             National Astronomical Observatory of Japan, Osawa 2-21-1, Mitaka, Tokyo 181-8588, Japan\\
             \email{doris.arzoumanian@nao.ac.jp}  
             }

   \date{Received ; accepted }

% \abstract{}{}{}{}{} 
% 5 {} token are mandatory
 
  \abstract
{Dense molecular filaments are widely believed to be representative of the initial conditions of star formation 
in interstellar clouds. Characterizing their physical properties such as their transverse size is therefore of paramount importance.
{\it Herschel} studies suggest that nearby ($d < 500$\,pc) molecular filaments have a typical half-power width $\sim$\,0.1pc, 
but this finding has been questioned recently on the ground that the measured widths tend to increase with the distance to the filaments.}
{Here we revisit the dependence of measured filament widths on distance or equivalently spatial resolution, 
in an effort to determine whether nearby molecular filaments have a characteristic half-power width 
or whether this is an artifact of the finite resolution of the {\it Herschel} data.}  
{We perform a convergence test on the well-documented B211/213 filament in Taurus, by degrading the resolution 
of the {\it Herschel} data several times and re-estimating the filament width from the resulting column density profiles. 
We also compare the widths measured for the Taurus filament and other filaments from the {\it Herschel} Gould Belt survey 
with those found for synthetic filaments with various types of simple, idealized column density profiles (Gaussian, power law, 
Plummer-like).
}
{We find that the measured filament widths do increase slightly as the spatial resolution worsens 
and/or the distance to the filaments increases. 
However, this trend is entirely consistent with what is expected from simple 
beam convolution for filaments with 
density profiles that are Plummer-like 
and have intrinsic half-power diameters $\sim\,$0.08--0.1\,pc 
and logarithmic 
slopes $1.5 < p < 2.5$ at large radii, as directly observed in many cases including the Taurus filament. 
Due to the presence of background noise fluctuations, deconvolution of the measured widths from 
the telescope beam is difficult and quickly becomes inaccurate. 
}    
{We conclude that the typical half-power filament width $\sim$\,0.1\,pc measured with {\it Herschel} in nearby clouds 
most likely reflects the presence of a true common scale in the filamentary structure of the cold interstellar medium, 
at least in the solar neighborhood. We suggest that this common scale may correspond to the magnetized turbulent correlation length 
in molecular clouds.}


   \maketitle
%
%________________________________________________________________

\section{Introduction}
\label{sec:intro}

The filamentary texture of the cold interstellar medium (ISM) is believed to play a central role in the star formation 
process 
\citep[e.g.,][for recent reviews]{Hacar+2022,Pineda+2022}. 
In particular, {\it Herschel} imaging surveys of nearby Galactic clouds \citep{Andre+2010,Molinari+2010,Hill+2011,Juvela+2012} 
have shown that filaments dominate the mass budget of molecular clouds at high densities \citep[e.g.,][]{Schisano+2014} 
and correspond to the birthplaces of most prestellar cores 
\citep[e.g.,][]{Konyves+2015,Marsh+2016,DiFrancesco+2020}.  
These findings support the notion that molecular filaments are representative of the initial conditions of the bulk of star 
formation in the Galaxy \citep[e.g.,][]{Andre+2014}. 
Characterizing their detailed physical properties is therefore of paramount importance. 
A key attribute of filamentary gas structures is their transverse diameter since the fragmentation properties of cylindrical filaments 
are expected to scale with the filament diameter, at least according to quasi-static fragmentation models 
\citep[e.g.,][]{Nagasawa1987,Inutsuka+1992}. 
Remarkably, 
{\it Herschel} observations suggest that nearby molecular filaments have a common 
width of $\sim$0.1\,pc, with some scatter around this 
value  \citep[][]{Arzoumanian+2011,Arzoumanian+2019}.
Indeed, analyzing the radial column density profiles of 599 {\it Herschel} filaments in 8 nearby clouds at $d < 500\,$pc, \citet{Arzoumanian+2019} 
found that the filaments of their sample share approximately the same half-power width $\sim$0.1\,pc with 
a spread of a factor of $\sim$2, 
regardless of their mass per unit length $M_{\rm line}$, 
whether they are 
{\it subcritical} with $M_{\rm line}   \lesssim 0.5\, M_{\rm line, crit}$,  
{\it transcritical} with  $0.5\, M_{\rm line, crit} \la M_{\rm line} \la 2\, M_{\rm line, crit}$, 
or  {\it thermally supercritical} with $M_{\rm line}  \ga 2\, M_{\rm line, crit}$, 
where $M_{\rm line, crit} = 2\, c_s^2/G $ is the thermal value of the critical mass per unit length \citep[e.g.,][]{Ostriker1964}, 
i.e., $\sim$$\, 16\, M_\odot \, {\rm pc}^{-1} $ 
for a sound speed $c_{\rm s} \sim 0.2$\,\kms 
or a gas temperature \hbox{$T \approx 10\,$K}. 
If confirmed, the existence of a typical filament width may have far-reaching implications 
as it introduces a characteristic length scale, and possibly a characteristic mass scale as well \citep[][]{Andre+2019},  
in the structure of molecular clouds believed to be hierarchical in nature and essentially scale-free.

However, the reliability 
of the filament widths derived from {\it Herschel} data has recently been questioned 
by \citet{Panopoulou+2022} 
who reported an intriguing correlation between the measured widths and the distances 
of the {\it Herschel} filaments in the \citet{Arzoumanian+2019} 
sample. 
\citet{Panopoulou+2022} 
suggested that the filament widths derived by \citet{Arzoumanian+2019}  
are strongly affected 
by the finite spatial resolution of the {\it Herschel} data and inconsistent with a characteristic intrinsic filament diameter $\sim 0.1$\,pc.

In this paper, 
we re-examine the effect of spatial resolution on the column density profiles derived from {\it Herschel} data 
and resulting estimates of filament widths. In Sect.~\ref{sec:taurus}, we perform a convergence test similar to that presented by \citet{Panopoulou+2022} 
for the Taurus B211/B213 filament. 
In Sect.~\ref{sec:models}, 
we perform the same convergence test on synthetic data for model filaments with simple 
radial density profiles (Gaussian, power law, Plummer-like). In Sect.~\ref{sec:hercshel}, the results of these tests 
are compared to the filament widths obtained by \citet{Arzoumanian+2019} and  \citet{Schisano+2014}.  
We find good agreement between all three sets of measured widths 
when the model filaments have Plummer-like 
density profiles with half-power diameters $\sim\,$0.08--0.1\,pc 
and logarithmic slopes $1.5 < p < 2.5$ at large radii. Our analysis therefore supports the conclusion that {\it Herschel} filaments tend  
to have similar column density profiles and a typical intrinsic half-power width $\sim\,$0.08--0.1\,pc. 
We conclude the paper in Sect.~\ref{sec:concl}, where we suggest that the typical filament width may be directly related  
to the correlation length of turbulent density and magnetic field fluctuations in molecular clouds. 

%__________________________________________________________________

\section{Convergence test for the Taurus filament}
\label{sec:taurus}

A detailed study of the radial (column) density profile of the B211/B213 filament in Taurus (at a distance $d \sim 140\,$pc)
was presented by  \citet{Palmeirim+2013} 
based on data from the {\it Herschel} Gould Belt survey (HGBS -- \citealp{Andre+2010}). 
Briefly, the mean transverse column density profile observed perpendicular to the filament is very well described 
by a Plummer-like model function of the form: 
\begin{equation}
N_p(r) = \frac{N_0}{[1 + \alpha_p\,(r/R_{\rm HP})^2]^{\frac{p-1}{2}}} + N_{\rm bg}, 
\end{equation}
\noindent
where $R_{\rm HP}$ is the half-power (HP) radius of the model profile,  
$p$ the power-law exponent of the corresponding density profile at radii much larger than $R_{\rm HP}$,
$\alpha_p = 2^{\frac{2}{p-1}}-1$, 
$N_0$ is the central column density, and $N_{\rm bg}$ the column density of the background cloud 
in the immediate vicinity of the filament. 
At the native half-power beam width (HPBW) resolution of the {\it Herschel} column density map used by \citet{Palmeirim+2013}, 
i.e., $18.2\arcsec $ or 0.012\,pc, the best-fit model profile has the following parameters: 
$D_{\rm HP}^{\rm int} = 2 \times R_{\rm HP} = 0.10 \pm 0.02$\,pc  
(taking beam convolution into account), $p = 2.0 \pm 0.4$, $N_0 = (1.5\pm0.2) \times 10^{22}\, \rm{cm}^{-2}$, 
and $N_{\rm bg} = (1.0\pm0.5) \times 10^{21}\, \rm{cm}^{-2}$.  
The HP 
diameter $D_{\rm HP}$ derived from fitting a Plummer model to the observed profile agrees well 
with the half-diameter $hd \equiv 2\,hr = 0.093$\,pc estimated in a simpler way, without any fitting,   
as twice the half-radius $hr$ where the background-subtracted column density profile drops to 
half of its maximum value on the filament crest \citep[][]{Arzoumanian+2019}. 
At the native {\it Herschel} resolution, the measured HP
width of the Taurus filament thus 
corresponds to 7.5--8.0 times the beam size. 

\begin{figure}        
\centerline{\resizebox{0.9\hsize}{!}{\includegraphics[angle=0]{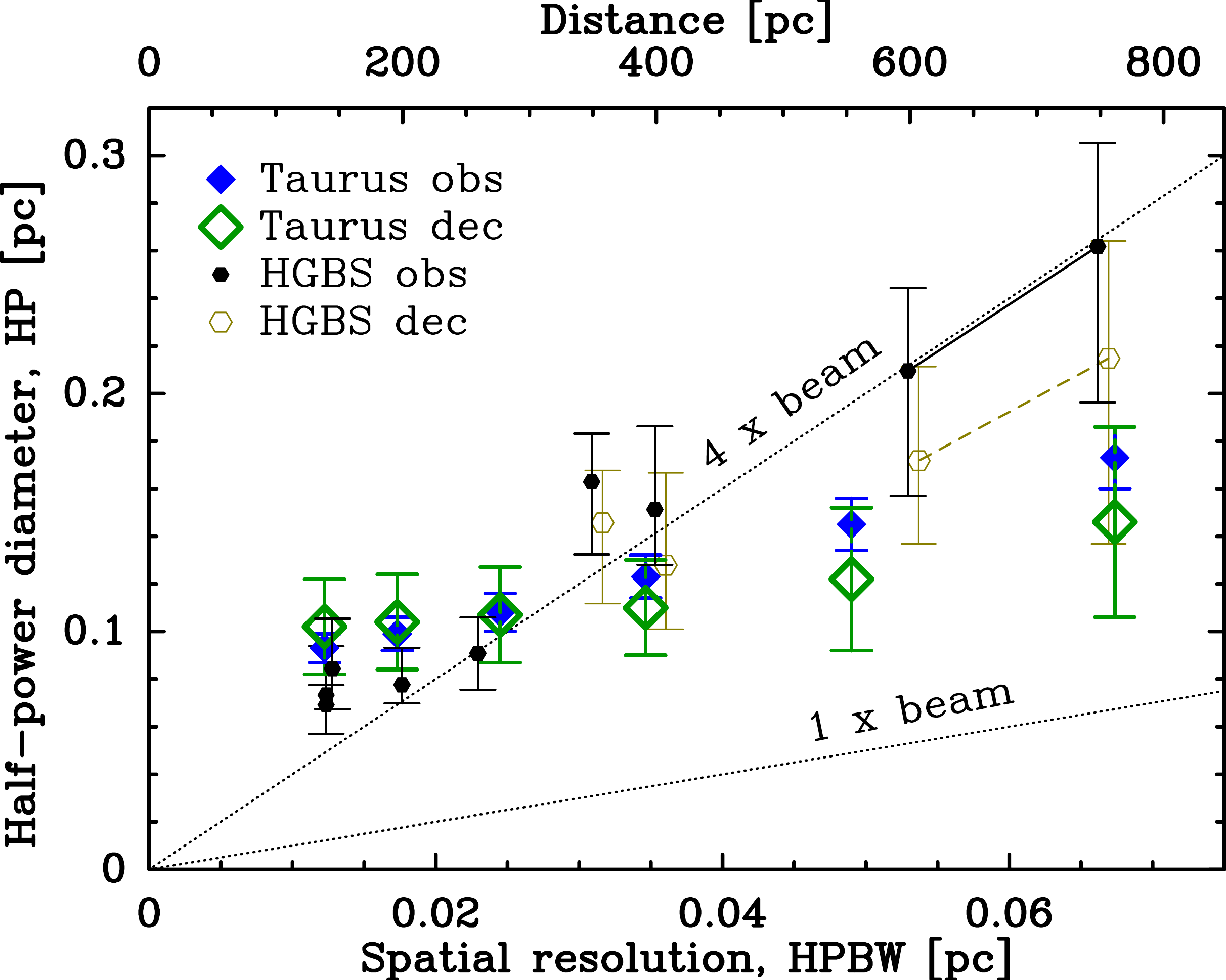}}}
\caption{Dependence of the measured half-power 
diameter HP$_{\rm obs}$ with spatial resolution (lower x-axis) 
or equivalently distance (upper x-axis) for the Taurus B211/3 
filament at several degraded resolutions given in 
Table~\ref{table_widths} 
(filled blue diamonds with error bars: simple $hd$ estimates; open green diamonds: 
deconvolved $D_{\rm HP}^{\rm dec}$ estimates) 
and other filaments from the {\it Herschel} GBS. 
{The black symbols and error bars show the median value and interquartile range of the filament widths 
measured in each of 8 HGBS clouds by \citet{Arzoumanian+2019}. 
The light brown symbols show deconvolved width estimates for the most distant clouds, 
based on the results of the Plummer tests of Sect.~\ref{sec:models} assuming $p=1.7$.}
{One cloud (IC5146) spans a significant range of distances according to {\it Gaia} data and is represented 
by two connected black/brown symbols.} 
}
\label{taurus_fil_width}
\end{figure}

To investigate the effect of spatial resolution on the column density profile and measured filament width, 
we performed a convergence test by degrading the resolution of the {\it Herschel} data, convolving the original column density map with progressively larger 
Gaussian kernels, and evaluating the filament width on the degraded maps 
in the same manner as on the original map. 
This is analogous to the convergence studies commonly performed on numerical simulations, where the resolution is progressively increased to check that the outcome 
eventually no longer depends  on numerical resolution 
\citep[cf.][]{Raskutti+2016}. 
The results of our test are provided in Table~\ref{table_widths} 
{(Appendix~\ref{App_tab})}
and displayed as blue and green 
symbols in Fig.~\ref{taurus_fil_width} 
as a function of both spatial resolution and equivalent distance (i.e., the distance at which 
the $18.2\arcsec $ angular resolution of the {\it Herschel} column density data matches the specified spatial resolution).  
A similar convergence test for the Taurus filament was presented by \citet{Panopoulou+2022}, 
although they emphasized the effect of cloud distance while we emphasize the effect of spatial resolution. 
Both effects are essentially equivalent but using spatial resolution instead of equivalent distance 
as the primary parameter is simpler and less confusing. 
({Note that} 
two figures in \citealp{Panopoulou+2022} 
were originally erroneous; see erratum of the paper.)

Figure~\ref{taurus_fil_width} also compares the results of the Taurus convergence experiment 
with the median apparent filament widths derived from {\it Herschel} data in the 8 nearby clouds
of the \cite{Arzoumanian+2019} sample (black dots). 
{Overall, the latter measurements are roughly consistent with the linear dependence 
HP$_{\rm obs} \approx 4 \times {\rm HPBW}$ \citep[cf.][]{Panopoulou+2022}, although 
the nearest HGBS clouds at $d \la 200\,$pc depart from this relation. 
In particular, it is clear from Fig.~\ref{taurus_fil_width} that the trend between 
HP$_{\rm obs}$ and HPBW for the Taurus filament (blue diamonds) 
is not consistent with a linear dependence with zero intercept.}  
In fact, extrapolating the width measurements to infinite resolution (i.e., ${\rm HPBW} = 0$) suggests 
that the intrinsic HP diameter of the Taurus filament is HP$_{\rm int} \approx 0.08$--0.1\,pc. 
Moreover, within the error bars, the Taurus B211/B213 filament provides a fairly good 
template of the behavior observed in the whole HGBS sample, 
{especially when deconvolved $D_{\rm HP}^{\rm dec}$ width estimates are compared}.

\section{Convergence curves for model filaments}
\label{sec:models}

To further clarify the influence of spatial resolution on real data, it is instructive to examine 
the effect for three simple types of cylindrical model filaments.

The simplest model is a filament with a Gaussian radial profile. 
In this case, the resolution effect due to convolution with the telescope beam 
(itself approximated by a Gaussian) is well known. 
The observed profile is also Gaussian with a full width at half maximum 
${\rm FWHM}_{\rm obs} = \sqrt{ {\rm FWHM}_{\rm int}^2 + {\rm HPBW}^2 }$, 
where FWHM$_{\rm int}  = {\rm HP}_{\rm int}$ is the intrinsic HP 
diameter of the filament profile 
and HPBW the half-power beam width. 
Provided that the data are sufficiently high signal to noise, the observations can easily be deconvolved from the telescope beam: 
{\hbox{${\rm FWHM}_{\rm dec} \equiv \sqrt{ {\rm FWHM}_{\rm obs}^2 - {\rm HPBW}^2 } = {\rm FWHM}_{\rm int}$}
(a formula referred to as naive deconvolution in the following).}
It is obvious from this simple example that, in the absence of any deconvolution, one expects a positive correlation between 
the measured HP widths ${\rm FWHM}_{\rm obs}$ and the spatial resolution of the data ${\rm HPBW}$. 

Another simple model is a filament with a power-law radial column density profile,  
$N_{\rm PL}(r) =  N_0 \, (r/r_0)^{-m}$, 
where $m = p-1$ and $p = m+1$ (with $m >0$) are the exponents of the power law for the column density and density profiles, respectively. 
In this case, the measured size ${\rm FWHM}_{\rm obs}$ (or HP$_{\rm obs}$)
of the filament profile is expected to directly scale with, and to be always slightly larger than, 
the beam size (see, e.g., \citealp{Adams1991} and \citealp{Ladd+1991} 
for spherical sources). 
To quantify the ${\rm FWHM}_{\rm obs}$ vs. ${\rm HPBW}$ correlation, we performed the convolution with a Gaussian beam numerically 
for a wide range of power-law exponents from $p \ga 1$ to $p = 3$. The results are displayed in Fig.~\ref{fig_pl}b 
{(Appendix~\ref{App0})}, 
which plots the ratio ${\rm FWHM}_{\rm obs}/{\rm HPBW} \equiv A_G(p)$ as a function of the index $p$ of the power-law profile. 
Observationally, both near-infrared extinction and submillimeter emission studies indicate that 
the column density profiles of dense molecular filaments have logarithmic slopes in the range $1.5 < p < 2.5$ at large radii  
\citep[e.g.,][]{Alves+1998,Lada+1999,Arzoumanian+2011,Arzoumanian+2019, Juvela+2012, Stutz+2016}.
For this range of $p$ exponents, Fig.~\ref{fig_pl}b shows that one expects the measured ${\rm FWHM}_{\rm obs}$ size to be from 
$\sim$5\% (for $p=2.5$) to  $\sim$90\% (for $p=1.5$) larger than the beam size if the filament profile is a pure power law.  

\begin{figure}               
\centerline{\resizebox{0.85\hsize}{!}{\includegraphics[angle=0]{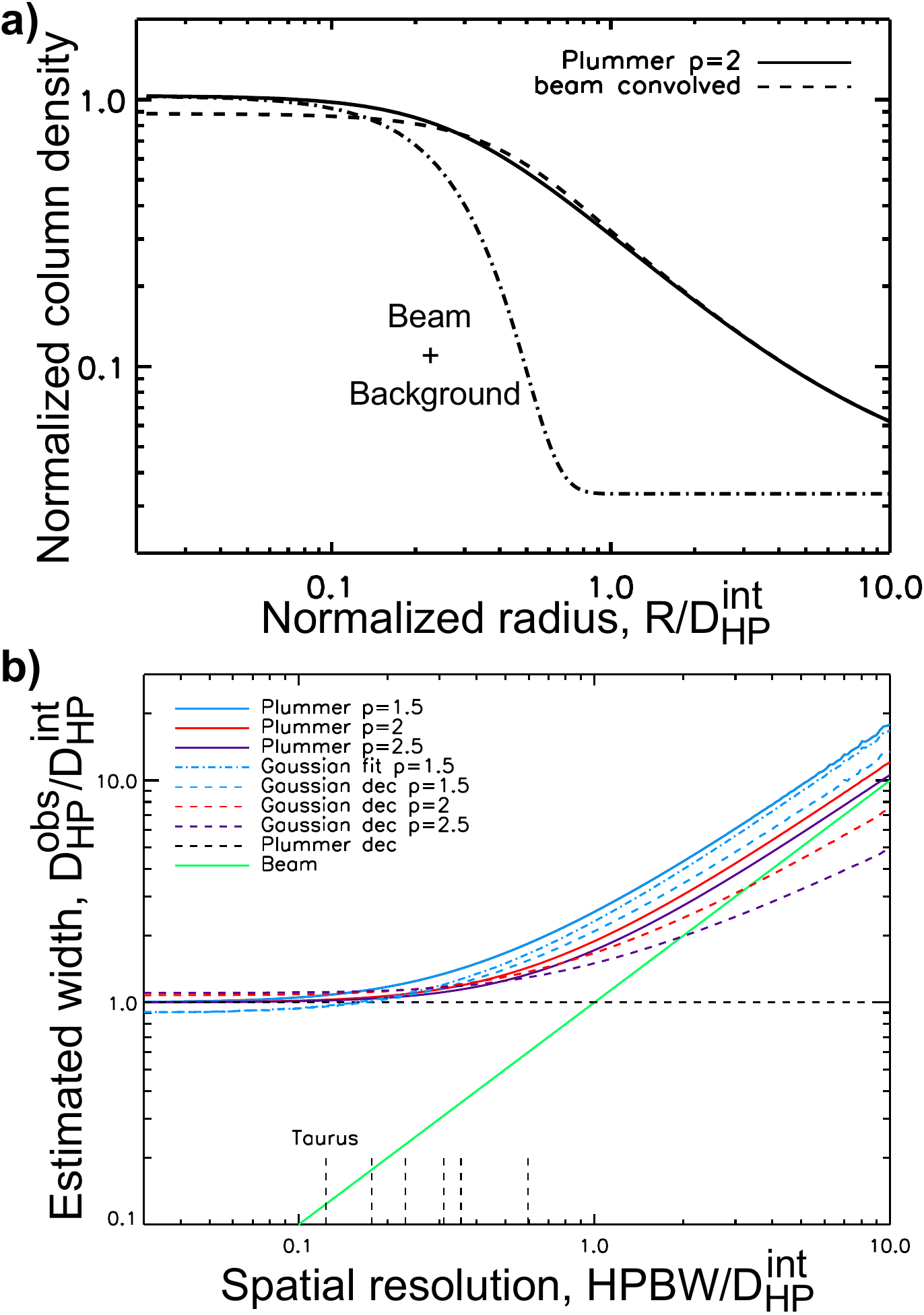}}}     
\caption{{\bf a)} 
Beam convolution effect for 
a model filament with a Plummer-like radial profile (see Eq.~(1)) 
with $p=2$. 
The solid curve shows the intrinsic radial profile on top of a uniform background, 
the dash-dotted curve 
the Gaussian beam (here assuming ${\rm HPBW}/D_{\rm HP}^{\rm int} = 1/2$) 
on top of the same background, and the dashed curve the 
filament profile after convolution with the beam.
{\bf b)} Measured HP diameter ($D_{\rm HP}$ or $hd$) against HPBW spatial resolution, both 
in units of the intrinsic HP 
diameter $D_{\rm HP}^{\rm int}$, for Plummer-like filaments with logarithmic density slopes $p=1.5$ (blue solid curve), $p=2$ (red solid curve), 
and $p=2.5$ (purple solid curve). 
{The blue dash-dotted curve shows the results of Gaussian fit measurements in the case of Plummer models with $p=1.5$.}
The horizontal dashed line shows the deconvolved HP 
diameter $D_{\rm HP}^{\rm dec}$, which coincides with $D_{\rm HP}^{\rm int}$ in the absence of noise. 
{In contrast, the dashed blue, red, and purple curves illustrate that naive deconvolution 
of Gaussian fit measurements fails to recover the intrinsic $D_{\rm HP}^{\rm int}$ diameters of Plummer models 
(with $p=1.5$, $p=2$, $p=2.5$, respectively), even in the absence of noise.}
The green line marks the beam HPBW. 
{The vertical dotted segments at the bottom mark the spatial resolutions of the 
\citet{Arzoumanian+2019} measurements assuming $D_{\rm HP}^{\rm int} = 0.1$\,pc.}
}
\label{fig_plummer}
\end{figure}

The third model we considered corresponds to a cylindrical filament with a Plummer-like radial column density profile 
as defined by Eq.~(1), which provides a much better fit to the profiles of star-forming filaments such as Taurus B211/B213 \citep[e.g.,][]{Palmeirim+2013}.   
We convolved this intrinsic model profile numerically for a wide range of Gaussian beam sizes 
(expressed in units of the intrinsic HP 
diameter $D_{\rm HP}^{\rm int}$) 
and several values of the power-law exponent $p$. We then fitted the model function of Eq.~(1) to the convolved profiles. 
The results are illustrated {by the solid curves} in Fig.~\ref{fig_plummer}b which plot the derived HP
widths $D_{\rm HP}^{\rm obs}$ 
as a function of the HP 
beam resolution ${\rm HPBW}$ 
for three relevant values of $p$ ($p=1.5$ in blue, $p=2$ in red, and $p=2.5$ in purple). 
Here,  $D_{\rm HP}^{\rm obs}$ corresponds either to a ``direct'' $hd$ estimate or to a $D_{\rm HP}$ estimate 
resulting from the Plummer-like fit in the terminology of Sect.~\ref{sec:taurus} and Table~\ref{table_widths}.
{(These two estimates are essentially indistinguishable.)}
{With some care \citep[cf. Sect.~3.3.3 of][]{Arzoumanian+2019}, reliable Gaussian fit estimates of 
the HP width of a Plummer-like profile can also be achieved. This is illustrated by the blue dash-dotted curve 
in Fig.~\ref{fig_plummer}b, which shows that Gaussian fit estimates are 
only slightly ($< 10\%$) lower than $D_{\rm HP}$ estimates in the $p=1.5$ case.
(For $p=2$ and $p=2.5$, the Gaussian results are even closer to the $D_{\rm HP}$ estimates.)}

\begin{figure}        
\centerline{\resizebox{0.85\hsize}{!}{\includegraphics[angle=0]{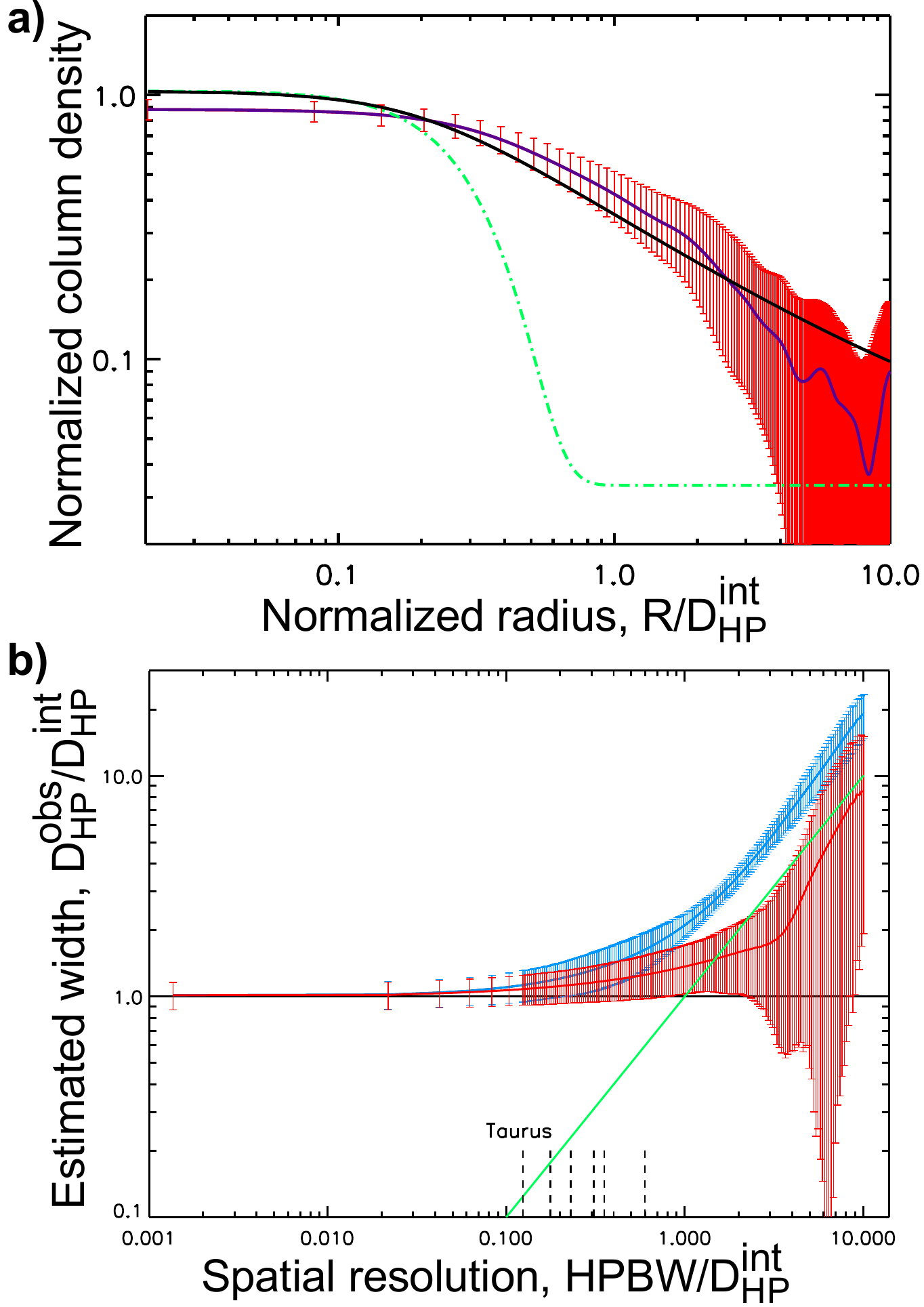}}}            
\caption{Similar to Fig.~\ref{fig_plummer} after adding realistic 
noise to a Plummer-like model filament with $p=1.7$. 
{\bf a)} Intrinsic 
radial profile (black solid curve) compared to the profile that would be 
observed 
{after convolution with the beam  (purple curve with red error bars) in the presence of typical background noise 
with a Kolmogorov-like power spectrum 
(see Appendix~\ref{App1})}. 
The green dash-dotted curve is the Gaussian beam, here assumed to be ${\rm HPBW} = D_{\rm HP}^{\rm int}/2$.
{\bf b)} Measured HP diameter against HPBW spatial resolution, 
prior to deconvolution ($hd$ or $D_{\rm HP}$, {blue curve with error bars}) 
and after deconvolution ($D_{\rm HP}^{\rm dec}$, {red curve with error bars}), 
for the model {including noise, whose one realization is shown in panel {\bf a)} 
(see Appendix~\ref{App1} for details)}. 
For comparison, the horizontal black 
line marks the intrinsic 
diameter $D_{\rm HP}^{\rm int}$ of the model,  
and the green line is the beam HPBW.  
The vertical dotted segments at the bottom mark the spatial resolutions of the 
\citet{Arzoumanian+2019} measurements assuming $D_{\rm HP}^{\rm int} = 0.1$\,pc for all HGBS filaments.
}
\label{fig_plummer_noise}
\end{figure}

The dependence of $D_{\rm HP}^{\rm obs}$  on ${\rm HPBW}$ for Plummer models
is qualitatively similar to the relation between ${\rm FWHM}_{\rm obs}$ and ${\rm HPBW}$ for Gaussian models 
at small beam sizes (\hbox{${\rm HPBW}/D_{\rm HP}^{\rm int} \la 1/3$})
and asymptotically approaches the linear relation \hbox{$D_{\rm HP}^{\rm obs} = A_G(p) \times {\rm HPBW}$} expected for filaments with 
power-law density profiles of index $p$ at large beam sizes  (${\rm HPBW}/D_{\rm HP}^{\rm int} \ga 1$) 
In the presence of negligible noise, it is possible to derive reliable deconvolved estimates 
of the HP diameter, denoted $D_{\rm HP}^{\rm dec}$, by fitting a model corresponding to Eq.~(1) convolved with a Gaussian beam 
to the data. 
This is illustrated by the black dashed line in Fig.~\ref{fig_plummer}b, which shows that the {Plummer deconvolution process} 
is perfect in the absence of noise. 
{In contrast, naively deconvolved Gaussian ${\rm FWHM}_{\rm dec}$ values (see above) 
largely overestimate the intrinsic $D_{\rm HP}^{\rm int}$ diameters of Plummer models when ${\rm HPBW}/D_{\rm HP}^{\rm int} \ga 1/3 $, 
even in the absence of noise (see dashed blue, red, and purple curves
in Fig.~\ref{fig_plummer}b).}
In the presence of a realistic level of 
background noise fluctuations (see Appendix~\ref{App1}), 
the Plummer deconvolution process 
{also quickly degrades} 
when ${\rm HPBW}/D_{\rm HP}^{\rm int} \ga 1/3 $ 
(see Fig.~\ref{fig_plummer_noise})  
and $D_{\rm HP}^{\rm dec}$ typically starts to increase with resolution 
(see {red curve and error bars} in Fig.~\ref{fig_plummer_noise}b). 
{$D_{\rm HP}^{\rm dec}$ estimates of the HP width nevertheless remain more accurate  
than naively deconvolved ${\rm FWHM}_{\rm dec}$ or $hd_{\rm dec}$ values.}
{Moreover, the curves of Fig.~\ref{fig_plummer}b can be inverted to provide average deconvolution factors
as a function of the observable  $D_{\rm HP}^{\rm obs}/{\rm HPBW}$ for Plummer profiles with any given $p$ index. 
This was used to derive the deconvolved estimates shown as light brown symbols in Fig.~\ref{taurus_fil_width}.} 
Comparison of the blue and green diamonds in Fig.~\ref{taurus_fil_width} with the blue and 
red curves in Fig.~\ref{fig_plummer_noise}b 
supports the conclusion that the Taurus B211/B213 filament 
has an intrinsic mean HP 
diameter $D_{\rm HP}^{\rm int} $ 
corresponding to $\sim$8 times the HPBW resolution of the {\it Herschel} column density map (see Sect.~\ref{sec:taurus} 
and \citealp{Palmeirim+2013}). 

\begin{figure}        
\centerline{\resizebox{0.95\hsize}{!}{\includegraphics[angle=0]{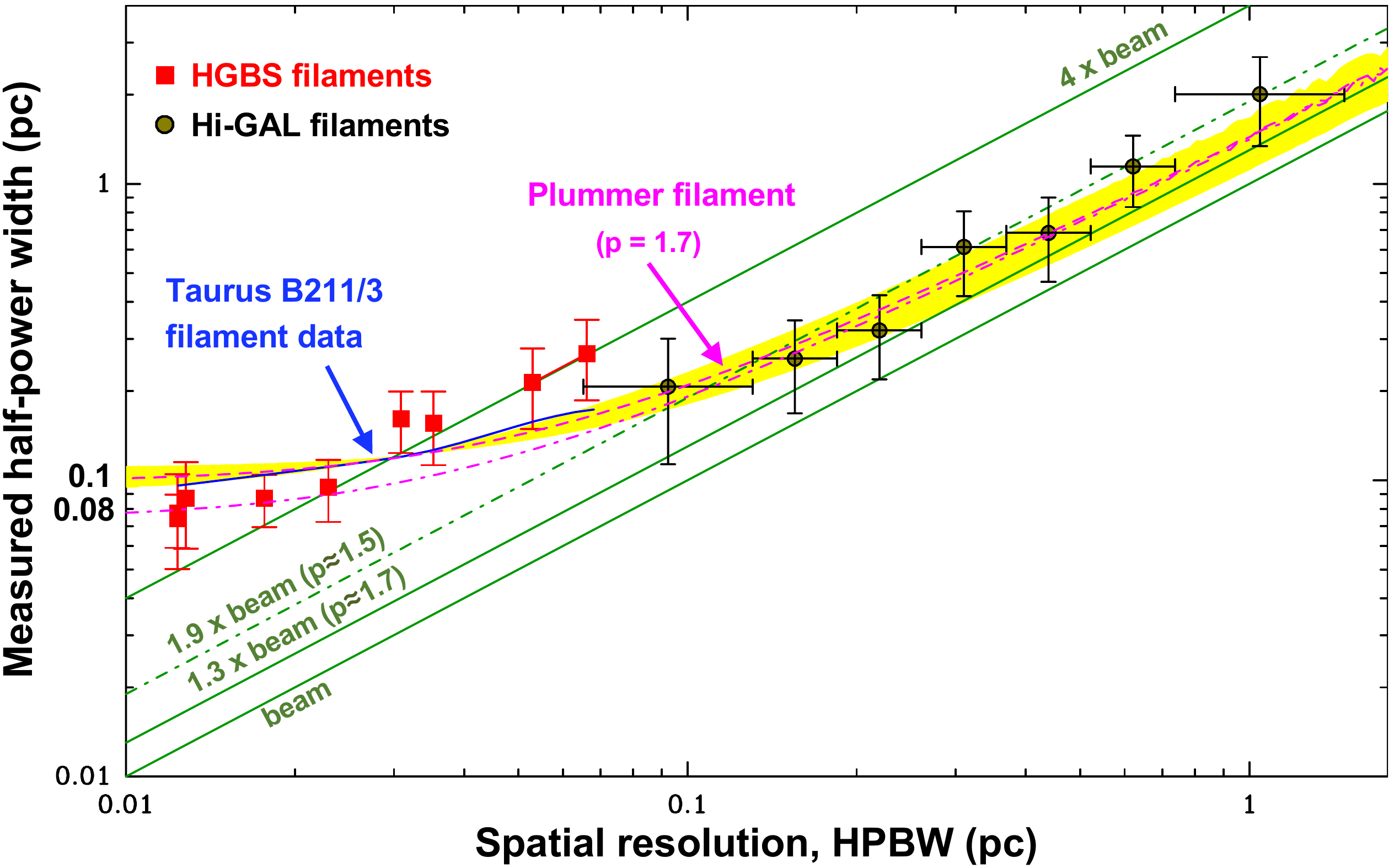}}} 
\caption{Mean 
{apparent}  
HP width vs. spatial resolution for 
both HGBS filaments (red squares from \citealp{Arzoumanian+2019}) 
and Hi-GAL filaments  (black circles;  
\citealp{Schisano+2014}). 
The vertical error bars correspond to $\pm$ the standard deviations of 
measured widths in each HGBS cloud or Hi-GAL resolution bin. 
{The horizontal error bars represent the Hi-GAL bin widths, which exceed the resolution uncertainties 
arising from the typical distance uncertainties.} 
The blue curve shows the results of the convergence test performed on the Taurus filament in Sect.~\ref{sec:taurus} 
(see blue diamonds in Fig.~\ref{taurus_fil_width}).  
The pink curves show the theoretical expectations for Plummer-like filaments 
with a logarithmic density slope $p = 1.7$ (consistent with the slope of the Taurus filament) and intrinsic HP 
widths of 0.1\,pc and 0.08\,pc, respectively (see Sect.~\ref{sec:models} and Fig.~\ref{fig_plummer}).
{In the 0.1\,pc case, the variations in expected HP width induced by variations of the 
$p$ index between 1.5 and 2.5 are displayed by a yellow shading.}
Green lines mark 
$1 \times {\rm beam}$, $1.3 \times {\rm beam}$,  $1.9 \times {\rm beam}$, 
and $4 \times {\rm beam}$.  
}
\label{fig_herschel}
\end{figure}

\section{Global comparison with Herschel filaments}
\label{sec:hercshel}

We may now confront the simple models of Sect.~\ref{sec:models} with 
the global results obtained on {\it Herschel} filaments. 
Figure~\ref{fig_herschel} shows the mean HP 
diameters found by \citet{Arzoumanian+2019} in 8 nearby HGBS clouds 
and by \citet{Schisano+2014} for a sample of Hi-GAL filaments
in the Galactic Plane,  
as a function of the HPBW spatial resolution 
of the {\it Herschel} data in each case.  
To generate this plot, the Schisano et al.  
sample was divided into 7 bins of 
spatial resolution between $\sim$\,0.1\,pc and $\sim$\,1\,pc (according to filament distance), 
and the mean filament width in each bin calculated from the results of \citet{Schisano+2014}.
First, it can be seen in Fig.~\ref{fig_herschel} that the whole set of {\it Herschel} width measurements is not consistent 
with a single linear relation such as $D_{\rm HP}^{\rm obs} = A(p) \times {\rm HPBW}$, 
expected for scale-free filaments. 
In particular, the {\it Herschel} data are inconsistent with the linear relation $D_{\rm HP}^{\rm obs} = 4 \times {\rm HPBW}$ 
advocated by \citet{Panopoulou+2022} for the HGBS filaments. 
The Hi-GAL filaments analyzed by \citet{Schisano+2014} have measured HP 
diameters that are typically $1.9 \pm 0.2$ times the HPBW spatial resolution (see green lines in Fig.~\ref{fig_herschel}). 
This is consistent with either pure power-law or unresolved Plummer-like density profiles 
with logarithmic slopes $p \sim 1.5$ at large radii. 
In contrast, the measured HP 
widths of nearby HGBS filaments 
are a factor $\sim$4 to $\sim$8 times the spatial resolution depending on filament distance, 
which is incompatible with pure power-law density profiles given the observed range 
of logarithmic slopes ($1.5 < p < 2.5$). 
A much lower ${\rm FWHM}_{\rm obs}/{\rm HPBW}$ ratio $\leq 1.9$ would indeed be expected in the latter case 
(see Sect.~\ref{sec:models} and Fig.~\ref{fig_pl}b). 
Quite remarkably, a single Plummer-like model with $D_{\rm HP}^{\rm int} \approx 0.08$--0.1\,pc and $p \approx 1.7$ 
(see pink curves in Fig.~\ref{fig_herschel}) 
can account for all of the {\it Herschel} measurements 
reasonably well. 
This suggests that {\it Herschel} filaments do have a typical HP 
width $\sim$0.08--0.1\,pc 
and typical power-law wings with $p \sim 1.5$--2, but that the flat inner 
portion of their 
column density profiles is unresolved by {\it Herschel} at the distances $d \sim 1$--3\, kpc 
of the Hi-GAL filaments in Fig.~\ref{fig_herschel}. 

\section{Concluding remarks}
\label{sec:concl}

{The convergence tests we performed on synthetic filaments 
indicate that filament width measurements are difficult to deconvolve from the telescope beam. 
In particular, naive Gaussian deconvolution of width measurements obtained on filaments with 
Plummer-like density profiles is ineffective (cf. Fig.~\ref{fig_plummer}b) and proper deconvolution assuming a Plummer model 
quickly becomes inaccurate in the presence of noise (Fig.~\ref{fig_plummer_noise}b). 
Given the logarithmic slopes $1.5 < p < 2.5$ of observed density profiles at large radii, 
our tests suggest that, without deconvolution, filament width measurements are reasonably accurate 
and overestimate the intrinsic HP widths  
of filaments by less than 10\%, 15\%, and 30\% on average when the apparent angular widths 
exceed the beam HPBW by a factor of $\sim$8, $\sim$6, and $\sim$4, respectively 
(cf. Fig.~\ref{fig_plummer}b). 
When the apparent filament width is only a factor of $\sim$2.5 higher than the HPBW, it may overestimate 
the intrinsic HP width by up to a factor of $\sim$2--3 and higher-resolution observations are desirable 
to improve the width estimates. 
When the apparent filament width is less than a factor of $\sim$2 broader than the HPBW, 
the measurements are dominated by the power-law wing of the filament profile and provide 
little information on the physical width of any flat inner region in the profile.
}

Overall, our analysis of the effects of finite spatial resolution on filament width measurements 
reinforces the conclusion 
of \citet{Arzoumanian+2011,Arzoumanian+2019} about the existence of a typical 
half-power filament diameter $\sim$0.1\,pc, at least in the case of nearby, high-contrast filamentary structures 
(i.e. with a contrast over the local background exceeding $\sim$50\%).  
{Our findings 
further emphasize 
the need for a robust theoretical explanation for the common filament width 
in  nearby molecular clouds.} 
While none of the current explanations is fully satisfactory from a theoretical point of view 
\citep[e.g.,][]{HennebelleInutsuka2019}, a promising 
interpretation is that the common filament width 
may be linked, at least initially, 
to the magneto-sonic scale below which interstellar turbulent flows become subsonic 
and incompressible (primarily solenoidal) in diffuse molecular gas \citep{Federrath2016, Federrath+2021}. 
The ubiquity of filamentary structures in both diffuse, non-self-gravitating molecular clouds such 
as the Polaris cirrus \citep[e.g.,][]{mamd+2010, Andre+2010} and numerical simulations of supersonic 
turbulence without gravity \citep[e.g.,][]{Padoan+2001, Pudritz+2013}  
suggests that dense molecular filaments somehow result from large-scale turbulent compression 
of interstellar material before gravity becomes important. 
Recently, \citet{Priestley+2022a,Priestley+2022b} argued that the observed distribution of filament widths 
can be explained if filaments are formed dynamically via converging, mildly supersonic flows
arising from large-scale turbulent motions.  
As pointed out by \citet{JaupartChabrier2021}, the sonic scale is directly related 
to the correlation length $l_c$ 
of initial density fluctuations generated by supersonic isothermal turbulence 
(before gravity starts to play a significant role). 
In other words, the typical filament width found in {\it Herschel} observations may correspond  
to the average size of the most correlated structures produced by turbulence in diffuse molecular clouds.
We stress that there is no contradiction between the existence of a finite correlation length for the (column) density field 
and the essentially scale-free power spectrum observed for column density fluctuations \citep[e.g.,][]{mamd+2010},  
since the correlation length $l_c \propto P(0)\,/$$\int P(\veck)\, d\veck$ only depends on the integral of the power spectrum $\int P(\veck)\, d\veck$ 
(i.e., the variance of the density field) and its value at zero spatial frequency $P(0)$ but not on the detailed shape 
of the power spectrum $P(\veck)$.  
Using column density data from the HGBS, \citet{JaupartChabrier2022} made rough estimates of the turbulent correlation length 
in the Polaris and Orion~B clouds 
and found values broadly consistent with $l_c \sim 0.1$\,pc, albeit with large uncertainties (up to a factor of $\sim$3--10).  
Following the method introduced by \citet{Houde+2009}, 
independent estimates of the magnetized turbulent correlation length have also been obtained from analyses 
of the angular dispersion function of polarization angles in recent high-quality dust polarimetry maps. 
For instance, using SOFIA HAWC$+$ data, \citet{Guerra+2021} found $l_c \sim 0.05$--0.15\,pc in 
most of the Orion OMC-1 field they observed and \citet{Li+2022} estimated $l_c \sim 0.075$--0.095\,pc 
in a significant portion of the Taurus B211 filament. 
One merit of interpreting the typical 
filament width as the turbulent correlation length 
is that it naturally accounts for the dispersion of 
width measurements around the mean 
$\sim$0.1\,pc value (both along the crest of each filament and from filament to filament), 
since the initial filament widths are only expected to 
match the sonic scale in a statistical sense.  
This interpretation is incomplete, however, because it does not explain how self-gravitating, thermally supercritical filaments 
maintain a roughly constant inner width while evolving. 
Further work is therefore clearly 
needed to provide a more complete explanation 
and clarify the role of the turbulent correlation length. 

\begin{acknowledgements}
We thank E. Schisano for kindly 
providing the width measurements from 
\citet{Schisano+2014}, used 
in Fig.~\ref{fig_herschel} of this paper. 
We acknowledge support from the French national programs of CNRS/INSU on stellar and ISM physics (PNPS and PCMI). 
PP acknowledges support from FCT/MCTES through Portuguese national funds (PIDDAC) by  grant UID/FIS/04434/2019 
and from fellowship SFRH/BPD/110176/2015 funded by FCT (Portugal) and POPH/FSE (EC).
This study has made use of data from the Herschel Gould Belt survey (HGBS) project, 
a Herschel Key Programme 
carried out by 
SPIRE Specialist Astronomy Group 3 (SAG 3), scientists of institutes in the 
PACS Consortium (CEA Saclay, INAF-IFSI Rome and INAF-Arcetri, KU Leuven, 
MPIA Heidelberg), and scientists of the Herschel Science Centre (HSC). 
\end{acknowledgements}


\bibliographystyle{aa}
\bibliography{filaments_ref}

\begin{appendix}

\section{Results of the convergence test on the Taurus filament}\label{App_tab}

{Here, we provide a table with the detailed results of the convergence test described in Sect.~\ref{sec:taurus}
on the Taurus B211/B213 filament:}

\begin{table}[!htbp]    
\caption{Estimated widths of the B211/B213 filament at various resolutions}
\label{table_widths}
\centering      % used for centering table
\begin{tabular}{c c c c c}   % centered columns (5 columns)
\hline\hline	% inserts double horizontal lines
Resol. & Resol. & $hd$ & $D_{\rm HP}$ & $D_{\rm HP}^{\rm dec}$  \\    	% table heading
 ($\arcsec$) & (pc) & (pc) & (pc) & (pc) \\
 (1) & (2) & (3) & (4) & (5) \\
\hline                        							% inserts single horizontal line
18.2 & 0.012 & $0.093 \pm 0.006$ & $0.11 \pm 0.01$ & $0.10 \pm 0.02$ \\		% inserting body of the table
25.7 & 0.018 & $0.099 \pm 0.007$ & $0.11 \pm 0.01$ & $0.10 \pm 0.02$ \\
36.4 & 0.025 & $0.108 \pm 0.008$ & $0.12 \pm 0.01$ & $0.11 \pm 0.02$ \\
51.5 & 0.035 & $0.123 \pm 0.009$ & $0.14 \pm 0.01$ & $0.11 \pm 0.02$ \\
72.8 & 0.050 & $0.145 \pm 0.011$ & $0.16 \pm 0.02$ & $0.12 \pm 0.03$ \\
99.0 & 0.067 & $0.173 \pm 0.013$ & $0.20 \pm 0.02$ & $0.15 \pm 0.04$ \\
\hline
\end{tabular}
\tablefoot{Col.~{\bf(3)}: Half-diameter $hd \equiv 2\,hr$ estimated without any fitting. Col.~{\bf(4)}: 
Half-power diameter $D_{\rm HP}$ derived by fitting Eq.~(1) to the observed mean column density profile (without any deconvolution). 
Col.~{\bf(5)}: Deconvolved half-power diameter $D_{\rm HP}^{\rm dec}$ derived by fitting a model corresponding to Eq.~(1) convolved 
with a Gaussian beam of HPBW given in Col.~{\bf(1)} (in $\arcsec$) and Col.~{\bf(2)} (in pc).  
}
\end{table}

\section{Tests on model filaments with power-law profiles}\label{App0}

\begin{figure}[!htbp]        
\centerline{\resizebox{0.85\hsize}{!}{\includegraphics[angle=0]{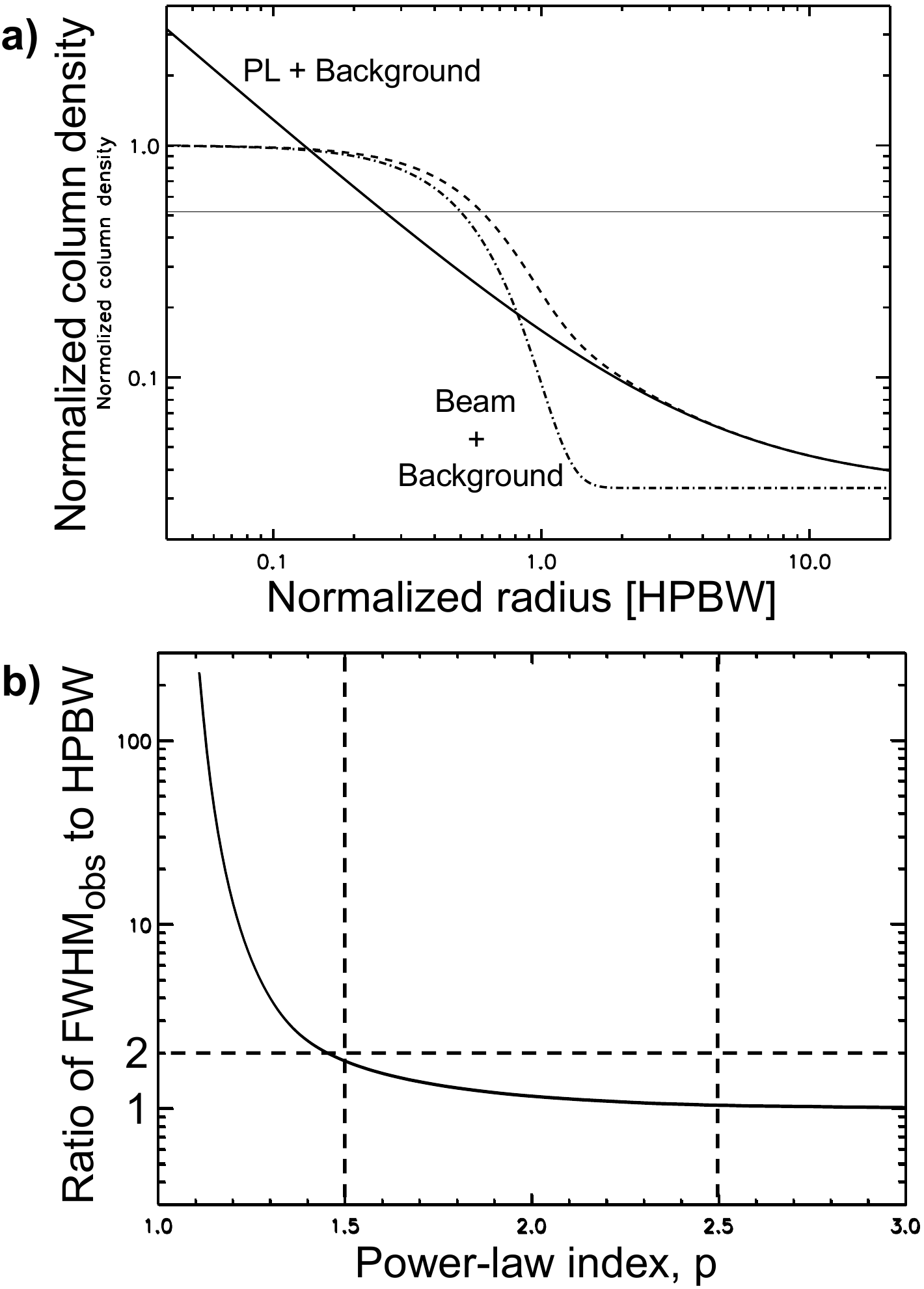}}}            
\caption{{\bf a)} 
Beam convolution effect 
in the case of a model filament with a power-law (PL) radial profile.
The solid curve shows the intrinsic PL 
profile on top of a uniform background,  
the dash-dotted curve 
the Gaussian 
beam on top of the same background, and the dashed curve the 
PL profile after convolution with the beam. 
The thin solid line marks the HP
level of both the convolved PL 
profile and the beam. 
{\bf b)} Plot of the ratio ${\rm FWHM}_{\rm obs}/{\rm HPBW} \equiv A_G(p)$ against the logarithmic slope of the radial density profile $p$ 
for PL filaments.
The two vertical dashed lines bracket the range of PL 
indices for observed filaments at large radii. 
}
\label{fig_pl}
\end{figure}

{Figure~\ref{fig_pl} illustrates and summarizes the results of the simple tests
we performed for model filaments with power-law column density profiles 
(see Sect.~\ref{sec:models}).}

\newpage
\section{Generating synthetic data for Plummer-like filaments with realistic background noise}\label{App1}

\begin{figure}[!htbp]        
\centerline{\resizebox{0.95\hsize}{!}{\includegraphics[angle=0]{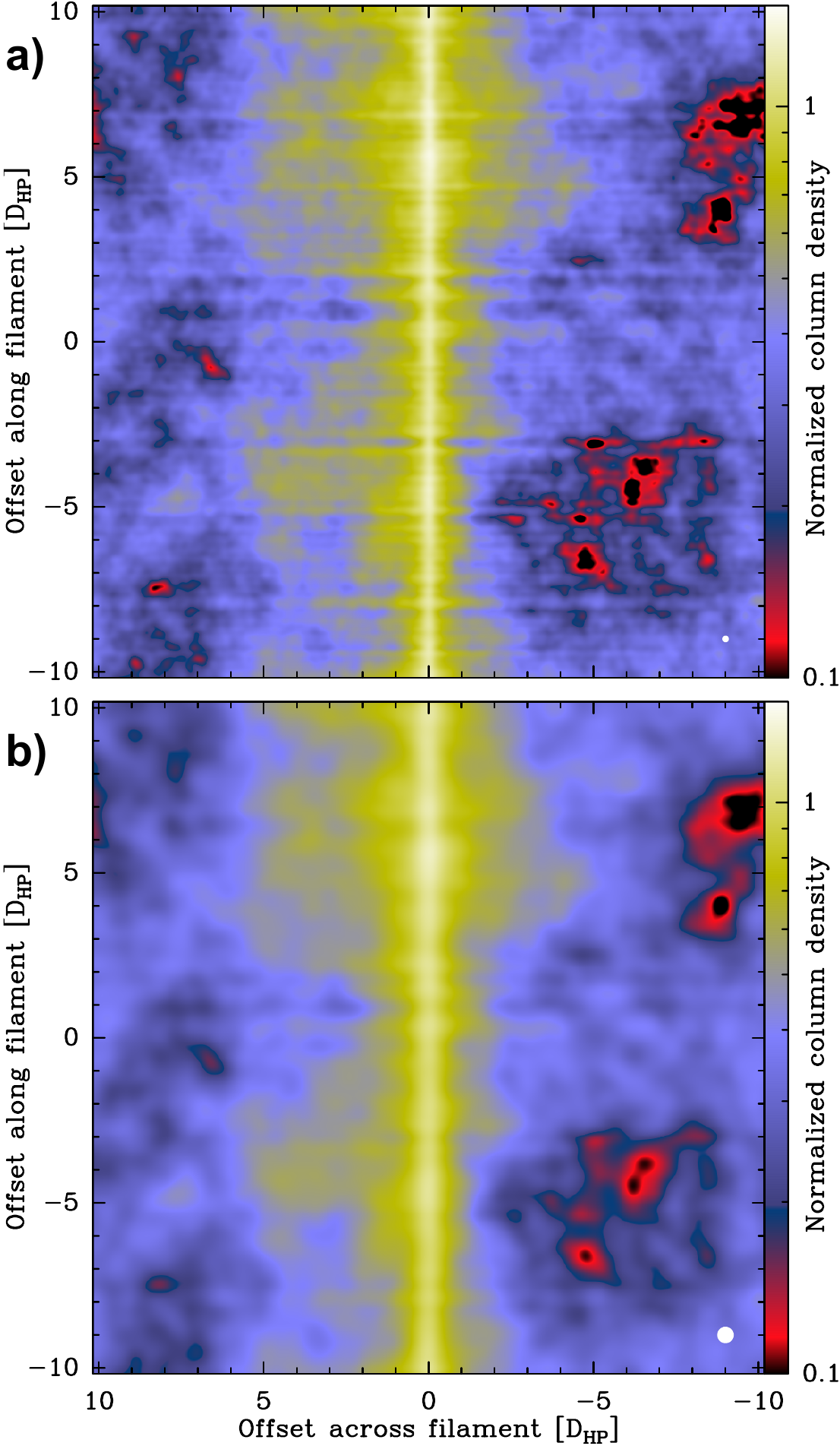}}} 
\caption{{Two examples of synthetic column density maps of a Plummer model filament with $p=1.7$ 
embedded in a structured background with a cirrus-like  $P(k) \propto k^{-2.7}$ power spectrum (see text).
{\bf a)} Model image ``observed'' at a HPBW resolution corresponding to $D_{\rm HP}^{\rm int}/10$. 
{\bf b)} Model image ``observed'' at a resolution ${\rm HPBW} = D_{\rm HP}^{\rm int}/2$, corresponding 
to the filament radial profile shown in Fig.~\ref{fig_plummer_noise}a. 
In both panels, positional offsets are given in units of the intrinsic HP width of the model filament, $ D_{\rm HP}^{\rm int}$.}}
\label{fig_maps_plummer}
\end{figure}

\noindent
{Here, we provide details on the method employed in Sect.~\ref{sec:models} to construct synthetic data 
corresponding to the observation of a cylindrical model filament with a Plummer-like column density profile (cf. Eq. 1) 
in the presence of realistic background ``noise'' fluctuations. 
The noise in {\it Herschel} maps of molecular clouds 
at wavelengths $\lambda \geq 160\, \mu$m is usually 
dominated by the ``structure noise'' induced by the fluctuations of the background cloud emission. 
Moreover, the power spectrum measured for far-infrared images of interstellar clouds is 
typically Kolmogorov-like with $P(k) \propto k^{-2.7}$ \citep[e.g.][]{mamd+2010}.
For the tests illustrated in Fig.~\ref{fig_plummer_noise}, 
we thus constructed a series of synthetic column density maps by adding randomly-generated maps of background fluctuations 
with such a $P(k) \propto k^{-2.7}$ power spectrum 
to the column density map 
of a Plummer model filament with a logarithmic density slope $p=1.7$ (assumed to lie in the plane of the sky). 
The standard deviation of the background noise map was fixed to 1/10 of the central column density of the model filament. 
We also introduced realistic random fluctuations of the intrinsic HP diameter 
of the model filament along its length; these random fluctuations had a lognormal distribution 
centered at $D_{\rm HP}^{\rm int} $ with a standard deviation 0.3~dex, as observed for HGBS filaments 
(see Fig.~7 of \citealp{Pineda+2022}). 
The synthetic maps were smoothed to a wide range of effective resolutions by convolving them with 
circular Gaussian kernels.
Two examples of such smoothed column density maps are displayed in Fig.~\ref{fig_maps_plummer}, 
including the map corresponding to the filament column density profile 
shown in Fig.~\ref{fig_plummer_noise}a observed at a resolution ${\rm HPBW} = D_{\rm HP}^{\rm int}/2$. 
For each synthetic column density map and each resolution, the distribution of resulting radial profiles along the model filament
was derived, which allowed us to construct a mean radial profile such as the one of Fig.~\ref{fig_plummer_noise}a, 
with error bars corresponding to the standard deviation of radial  
profiles at each radius. 
The model represented by Eq.~(1), both before and after convolution with a Gaussian beam, 
was then fitted to each of these mean radial profiles 
to provide $D_{\rm HP}^{\rm obs}$ and $D_{\rm HP}^{\rm dec}$ estimates of the filament HP width 
at each resolution.
A total of 1000 realizations of the random background fluctuations were generated to construct the
plot shown in Fig.~\ref{fig_plummer_noise}b, where the blue and red solid curves represent, 
as a function of HPBW resolution, the mean $D_{\rm HP}^{\rm obs}$ and $D_{\rm HP}^{\rm dec}$ values averaged over the 1000 realizations, 
and the error bars correspond to the standard deviations measured for the distributions of values at each resolution.
}

\section{Distribution of {\it Herschel} filament widths}\label{App2}

\begin{figure}[!htp]        
\centerline{\resizebox{0.95\hsize}{!}{\includegraphics[angle=0]{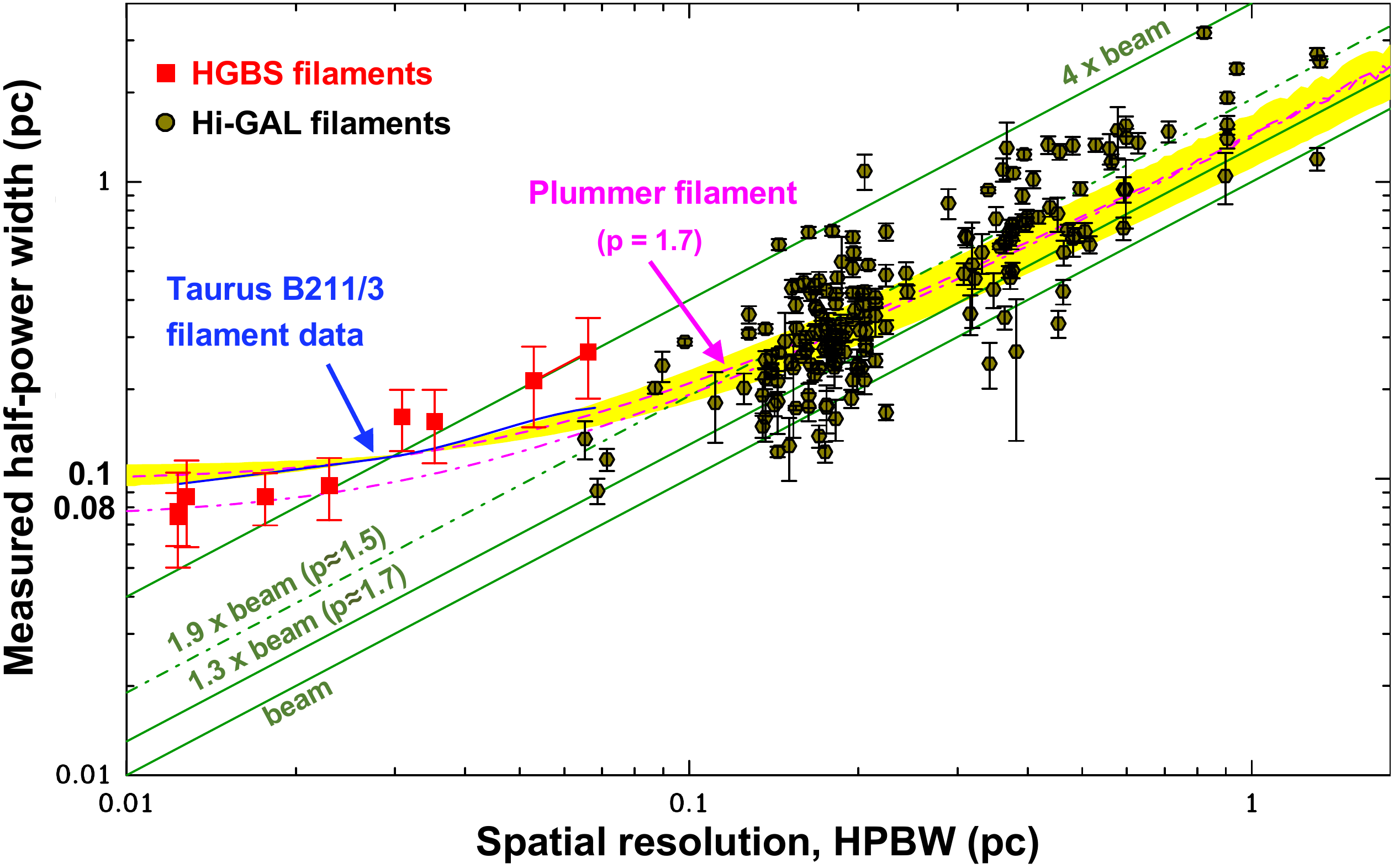}}} 
\caption{{Mean HP width versus spatial resolution for 
both HGBS filaments (red squares from \citealp{Arzoumanian+2019}) 
and Hi-GAL filaments  (black circles from \citealp{Schisano+2014}). 
The red error bars correspond to $\pm$ the standard deviations of 
measured widths in each HGBS cloud. 
The black error bars correspond to  $\pm$ the standard deviation of individual 
width estimates along each Hi-GAL filament in the Schisano et al. sample. 
The lines, curves, and yellow shading are the same as in Fig.~\ref{fig_herschel}.}
}
\label{fig_herschel_supp}
\end{figure}

{In this appendix, we show a plot of half-power width against spatial resolution similar to Fig.~\ref{fig_herschel},
but where we display all 
individual width measurements from \citet{Schisano+2014}
instead of binning the distribution of Hi-GAL widths (Fig.~\ref{fig_herschel_supp}). In this way, the full scatter of Hi-GAL width measurements 
may be better appreciated. 
Like in Fig.~\ref{taurus_fil_width} and Fig.~\ref{fig_herschel}, 
the distances adopted in Fig.~\ref{fig_herschel_supp} for the HGBS clouds 
are those indicated by {\it Gaia} data \citep[cf.][]{Panopoulou+2022}. 
The HGBS cloud with the most uncertain distance (IC5146) is represented by two connected symbols.  
The distances adopted for the Hi-GAL filaments are those used by \citet{Schisano+2014}, whose 
typical uncertainties are smaller than the horizontal error bars in Fig.~\ref{fig_herschel}. 
}

\newpage
\section{Effect of a non-Gaussian beam}\label{App3}

\begin{figure} [!hbp]             
\centerline{\resizebox{0.85\hsize}{!}{\includegraphics[angle=0]{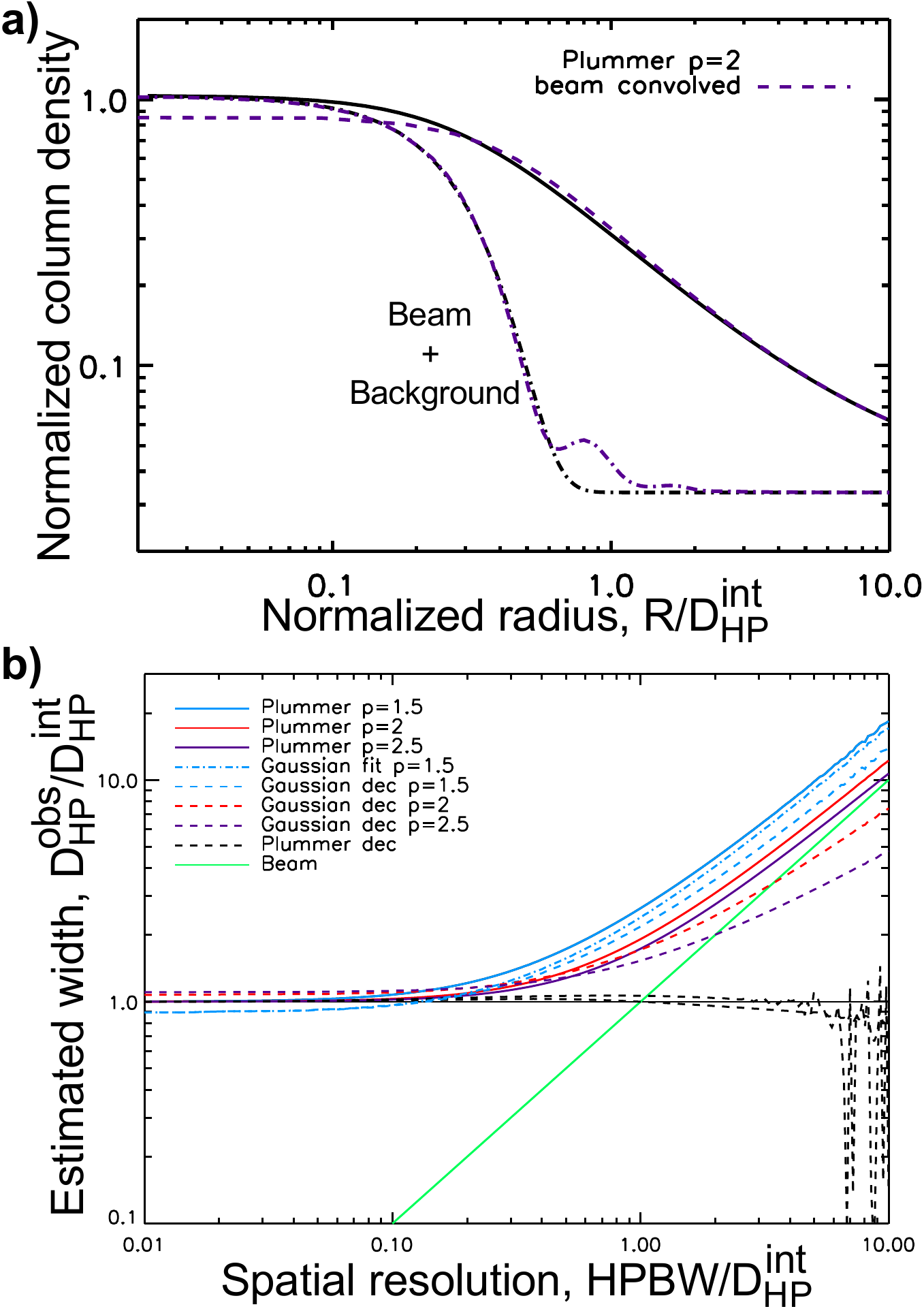}}}     
\caption{{Similar to Fig.~\ref{fig_plummer} but assuming that the effective beam corresponds 
to the real SPIRE 250\,$\mu$m beam as opposed to a strictly Gaussian beam.}
{\bf a)} 
{Convolution of a model filament with a $p=2$ Plummer-like radial profile 
by a non-Gaussian beam corresponding to the SPIRE beam at 250\,$\mu$m.
The purple dash-dotted curve shows this non-Gaussian beam on top of the same 
uniform background as in Fig.~\ref{fig_plummer}a. 
The purple dashed curve is the filament profile after convolution with this beam 
(assuming the same ${\rm HPBW}/D_{\rm HP}^{\rm int} = 1/2$ ratio as in Fig.~\ref{fig_plummer}a). 
The black solid and dash-dotted curves are the same as in Fig.~\ref{fig_plummer}a.}
{\bf b)} {Same as in Fig.~\ref{fig_plummer}b when the convolved synthetic data 
are constructed with a non-Gaussian beam of shape as shown in panel {\bf a} 
but the estimated $D_{\rm HP}^{\rm dec}$ diameters are derived as in 
Fig.~\ref{fig_plummer}b assuming a strictly Gaussian beam. 
}
}
\label{fig_plummer_spire}
\end{figure}

{The synthetic data discussed in Sect.~\ref{sec:models} were smoothed 
to various effective resolutions assuming a strictly Gaussian beam.  
In reality, the {\it Herschel} beams are not strictly Gaussian and 
include low-level non-Gaussian features \citep[e.g.,][]{Aniano+2011}. 
To assess the potential impact of these non-Gaussian features 
on the tests performed in Sect.~\ref{sec:models}, we repeated the numerical experiment  
summarized in Fig.~\ref{fig_plummer} assuming that the effective beam of {\it Herschel} column 
density maps had a shape similar to that of the true {\it Herschel}/SPIRE beam at 250\,$\mu$m 
at all resolutions (see purple dash-dotted curve in Fig.~\ref{fig_plummer_spire}a). 
As shown in Fig.~\ref{fig_plummer_spire}b, the curves of derived HP widths (Plummer $D_{\rm HP}^{\rm obs}$ 
and Gaussian FWHM$_{\rm obs}$) as a function of HP beam resolution ${\rm HPBW}$ 
are almost identical to those in Fig.~\ref{fig_plummer}b. 
This indicates that the non-Gaussian structure of the {\it Herschel} beams has very little influence 
on the results presented in Sect.~\ref{sec:models}. 
The only visible effect is that the Plummer deconvolution process 
and resulting $D_{\rm HP}^{\rm dec}$ estimates of the filament width (see black dashed lines in Fig.~\ref{fig_plummer_spire}b)
are no longer perfectly accurate, even in the absence of noise.
This is hardly surprising since accurate beam deconvolution requires both high signal to noise 
and excellent knowledge of the beam shape. 
}

\end{appendix}

\end{document}